\documentclass[12pt,twoside]{article}

\evensidemargin=\oddsidemargin
\def\Title#1#2#3{%
    \baselineskip=18pt
    \begin{center}
          {\large\bf\uppercase{#1} \\ }
          \bigskip\bigskip
          {#2} \\
          {#3} \\
    \end{center}}
\long\def\Abstract#1{%
         \bigskip
         \parbox{0.93\textwidth}{%
                 \begin{center}
                       {\bf Abstract} \\
                 \end{center}
                 \medskip{\baselineskip=14pt #1}
                 \vss}
         \bigskip}

\makeatletter
\renewcommand{\section}%
 {\@startsection{section}{1}{0pt}%
  {-3.25ex plus -1ex minus -.2ex}{1.5ex plus .2ex}%
  {\vspace*{5mm}\raggedright\large\bf }}
\renewcommand{\thesection}{\arabic{section}.}
\@addtoreset{equation}{section}
\renewcommand{\@eqnnum}{(\thesection\theequation)}
\renewcommand{\p@equation}{\thesection}
\makeatother

\begin{document}

\vspace*{1cm}

\Title{THE PROSPECTS \\
FOR EXTENDED PHASE SPACE APPROACH \\
TO QUANTIZATION OF GRAVITY}%
{T. P. Shestakova}%
{Department of Theoretical and Computational Physics, Rostov State University, \\
Sorge St. 5, Rostov-on-Don 344090, Russia \\
E-mail: {\tt shestakova@phys.rsu.ru}}

\Abstract{A brief review of main features of the new approach named
``quantum geometrodynamics in extended phase space'' is given and
its possible prospects are discussed. Gauge degrees of freedom are
treated as a subsystem of the Universe which affects the evolution
of the physical subsystem. Three points can be singled out when
the gauge subsystem shows itself as a real constituent of the
Universe: a chosen gauge condition determines the form of equation
for the physical part of wave function, the form of density matrix
and the measure in physical subspace. An example is considered when
a physically relevant choice of gauge condition leads to almost
diagonal density matrix. The analogy between a transition to
another reference frame (another basis in physical subspace) and a
transition to accelerating reference frame in Rindler space is
suggested.}

\section{Introduction}
My talk is devoted to a new approach which was proposed by G. M.
Vereshkov, V. A. Savchenko and me \cite{SSV1,SSV2,SSV3,SSV4} and
named ``quantum geometrodynamics in extended phase space''. The
results of our work have already been reported at several
conferences (see, for example, \cite{Shest2,Shest3}), in particular,
at ``COSMION-99'' \cite{Shest1}. So the goal of this my talk is to
give a review of main features of the new approach and to discuss
its possible prospects.

A search for a new approach was inspired by the well-known problems
of the Wheeler -- DeWitt quantum geometrodynamics such as the
problem of time and the problem of reparametrization noninvariance.
It is a generally known fact that, while at the classical level
different forms of gravitational constraints corresponding to
various choice (parametrizations) of gauge variables are
equivalent, at the quantum level they lead to non-equivalent
conditions on a wave function. One can find in the literature
different solutions to the Wheeler -- DeWitt equation which
correspond to different parametrizations of a gauge variable and
are not reducible to each other \cite{HH,Vil,Hal1,GL,CFL,Fil,DF}. A
choice of the gauge variable (as a rule, it is the lapse function
$N$) is inevitable and, as was shown in our papers \cite{SSV4,
Shest1}, implies some choice of a reference frame. Indeed, the
choice of gauge variables together with imposing any additional
conditions on these variables define equations for $g_{0\mu}$
components of the metric which fix a reference frame according to
Landau and Lifshitz. One could consider the condition of
independence of $N$ on physical degrees of freedom \cite{HP} as
such an additional condition on the gauge variable, but even in
this case it would imply the choice of a certain reference frame.
It becomes especially important in case of simple cosmological
models when we deal with the only gauge degree of freedom $N$. It
has been emphasized by some authors that the (3+1)-slicing in the
Arnowitt -- Deser -- Misner formalism is equivalent to some kind of
gauge fixing and there is the inconsistency between the
(3+1)-slicing of a global manifold and the requirement for a wave
function to be invariant under space diffeomorphisms and time
displacement \cite{MM1,MM2,MM3}. All the above leads to the
conclusion that we failed to construct quantum geometrodynamics
which would be gauge invariant in a strict mathematical sense.

It may seem that these mathematical problems are typical for
canonical approach developed by Dirac for constrained systems and
the situation would be different if one turned to a more powerful
path integral quantization of gauge theories of Batalin, Fradkin
and Vilkovisky (BFV). As a matter of fact, in the path integral
approach one faces the same problems in another appearance. The BFV
effective action is constructed in such a way that dynamics in
extended phase space (EPS) turns out to be equivalent the Dirac
generalized Hamiltonian dynamics. The main role is given to
gravitational constraints. The Wheeler -- DeWitt equation can be
obtained in this approach as a consequence of the requirement of
the BRST invariance of a state vector, $\Omega\,|\Psi\rangle=0$
(for simple cosmological models $\Omega$ is a linear combination of
the constraints). On the other hand, the Wheeler -- DeWitt equation
can be derived from the BFV path integral under the assumption of
its BRST invariance which is ensured by asymptotic boundary
conditions on ghosts and Lagrange multipliers (see \cite{Hal1}).
So, the Wheeler -- DeWitt equation, even if derived from the path
integral by formal procedure, inherits all the problems we have met
in canonical quantization: the inevitability of choice of
parametrization, etc. Moreover, one should impose asymptotic
boundary conditions which are justified only in asymptotically flat
spacetime, while already in case of a close universe or a universe
with another non-trivial topology we have no grounds to impose
asymptotic boundary conditions.

The next point which I would like to mention is the difference
between the group of transformations generated by gravitational
constrains and that of gauge transformations of the Einstein
theory. In fact, already at the classical level we deal with two
different theories of gravity which are invariant under different
groups of transformations in Lagrangian and Hamiltonian
formulations. In the path integral quantization these
transformations define the structure of ghost sectors which also
appear to be different. The two formulations could enter into
agreement only in a gauge-invariant sector which can be singled out
by asymptotic boundary conditions; the later ones must supposedly
pick out trivial solutions for ghosts and Lagrange multipliers to
ensure gauge-invariant dynamics. However, if one consider the
Universe as a system which, in general, does not possess asymptotic
states, one have to pose the question: what formalism should one
prefer? What are consequences of the fact that we consider the path
integral without asymptotic boundary conditions? What will be a
role of gauge degrees of freedom, which were traditionally
considered as redundant, in this new approach?

\section{Hamiltonian dynamics in EPS: our approach}
It is more convenient to illustrate our approach for a simple
minisuperspace model with a gauged action
\begin{equation}
\label{action}
S=\!\int\!dt\,\biggl\{
  \displaystyle\frac12 v(\mu, Q)\gamma_{ab}\dot{Q}^a\dot{Q}^b
  -\frac1{v(\mu, Q)}U(Q)
  +\pi_0\left(\dot\mu-f_{,a}\dot{Q}^a\right)
  -i w(\mu, Q)\dot{\bar\theta}\dot\theta\biggr\}.
\end{equation}
Here $Q=\{Q^a\}$ stands for physical variables such as a scale
factor or gravitational-wave degrees of freedom and material
fields, and we use an arbitrary parametrization of a gauge variable
$\mu$ determined by the function $v(\mu, Q)$. In the case of
isotropic universe or the Bianchi IX model $\mu$ is bound to the
scale factor $a$ and the lapse function $N$ by the relation
\begin{equation}
\label{paramet}
\displaystyle\frac{a^3}{N}=v(\mu, Q).
\end{equation}
\begin{equation}
\label{w_def}
w(\mu, Q)=\frac{v(\mu, Q)}{v_{,\mu}};\quad
v_{,\mu}\stackrel{def}{=}\frac{\partial v}{\partial\mu}.
\end{equation}
$\theta,\,\bar\theta$ are the Faddeev -- Popov ghosts after
replacement $\bar\theta\to -i\bar\theta$.

In the class of gauges not depending on time
\begin{equation}
\label{frame_A}
\mu=f(Q)+k;\quad
k={\rm const},
\end{equation}
which can be presented in a differential form,
\begin{equation}
\label{diff_form}
\dot{\mu}=f_{,a}\dot{Q}^a,\quad
f_{,a}\stackrel{def}{=}\frac{\partial f}{\partial Q^a}.
\end{equation}
the Hamiltonian can be obtained in a usual way, according to the
rule $H=P\dot Q-L$, where $(P,Q)$ are the canonical pairs of
extended phase space, by introducing momenta conjugate to all
degrees of freedom including the gauge and ghost ones,
\begin{equation}
\label{Hamilt}
H=P_a\dot Q^a+\pi\dot\mu+\bar\rho\dot\theta+\dot{\bar\theta}\rho-L
 =\frac12G^{\alpha\beta}P_{\alpha}P_{\beta} +\frac1{v(\mu,Q)}U(Q)
 -\frac i{w(\mu,Q)}\bar\rho\rho,\nonumber\\
\end{equation}
where $\alpha=(0,a),\;Q^0=\mu$,
\begin{equation}
\label{Galphabeta}
G^{\alpha\beta}=\frac1{v(\mu,Q)}\left(
\begin{array}{cc}
f_{,a}f^{,a}&f^{,a}\\
f^{,a}&\gamma^{ab}
\end{array}
\right).
\end{equation}
Varying the effective action (\ref{action}) with respect to
$Q^a$,$\mu$,$\pi$ and $\theta$,$\bar\theta$ one gets,
correspondingly, motion equations for physical variables, the
constraint, the gauge condition and equations for ghosts.
The extended set of Lagrangian equations is complete in
the sense that it enables one to formulate the Cauchy problem. The
explicit substitution of trivial solutions for ghosts and the
Lagrangian multiplier $\pi$ to this set of equations turns one
back to the gauge-invariant classical Einstein equations.

It is not difficult to check that the system of Hamiltonian
equations in EPS
\begin{equation}
\label{Hamilt.Eqs.}
\dot{P} = - \frac{\partial H}{\partial Q};\quad
\dot{Q} = \frac{\partial H}{\partial P}
\end{equation}
is completely equivalent to the extended set of Lagrangian
equations, the constraint and the gauge condition acquiring the
status of Hamiltonian equations. The idea of extended phase space
is exploited in the sense that gauge and ghost degrees of freedom
are treated on an equal basis with other variables. This gave rise
to the name ``quantum geometrodynamics in extended phase space''.

Let us emphasize that the Hamiltonian dynamics is constructed here
in a different way than in the BFV approach. For example, the
Hamiltonian constraint is modified and looks like follows:
\begin{equation}
\label{pi-Ham}
\dot\pi=\frac{1}{v^2(\mu,Q)}v_{,\mu}
 \left[\frac12\left(P_aP^a+2\pi f_{,a}P^a+\pi^2f_{,a}f^{,a}\right)
  +U(Q)\right]-\frac i{w^2(\mu,Q)}w_{,\mu}\bar\rho\rho.
\end{equation}
The gauge-dependent terms can be eliminated making use of trivial
solutions for $\pi$ and ghosts. Together with the restriction on
the class of admissible parametrization,
$v(\mu,Q)=\frac{u(Q)}{\mu}$, it reduces the constraint to the form
\begin{equation}
\label{Dir-constr}
{\cal T}=\frac1{2u(Q)}P_aP^a+\frac1{u(Q)}U(Q)=0.
\end{equation}
Thus, the Hamiltonian constraint (\ref{Dir-constr}) can be restored
by means of asymptotic boundary conditions. We come to the
conclusion that the Dirac primary and secondary constraints
$\pi=0,\quad {\cal T}=0$ correspond to a particular situation when
it is possible to single out the trivial solutions for $\pi$ and
ghosts by asymptotic boundary conditions. In this sense, both the
Dirac and the BFV quantization schemes are applicable to systems
with asymptotic states only.

Since the Hamiltonian dynamics in EPS is completely equivalent to
Lagrangian dynamics, the group of transformations in EPS
corresponds to the group of gauge transformations in the
Lagrangian formalism. One can construct the BRST generator,
\begin{equation}
\label{BRSTgen}
\Omega
 =w(Q,\mu)\;\pi\dot\theta-H\theta
 =-\;i\;\pi\rho-H\theta.
\end{equation}
It is easy to check that (\ref{BRSTgen}) generates transformations
in EPS which are identical to the BRST transformations in the Lagrangian
formalism. On the other hand, the generator (\ref{BRSTgen}) does
not coincide with the one constructed according to prescriptions
by BFV which for the present model has an especially simple form:
\begin{equation}
\label{Omega_BFV}
\Omega_{BFV}=\eta^{\alpha}{\cal G}_{\alpha}={\cal
T}\theta-i\pi\rho,
\end{equation}
where ${\cal G}_{\alpha}=(\pi,\;{\cal T})$ is the full set of
constraints. As was mentioned above, the Wheeler -- DeWitt equation
${\cal T}\,|\Psi\rangle=0$ immediately follows from the requirement
of BRST invariance $\Omega_{BFV}\,|\Psi\rangle=0$ due to
arbitrariness of BFV ghosts $\{\eta^{\alpha}\}$.

Because of the difference in groups of transformations the BFV
charge (\ref{Omega_BFV}) turns our to be irrelevant in this
consideration. At the same time, the ``new'' BRST generator
(\ref{BRSTgen}) cannot be presented as a combination of constraints
and does not lead to the Wheeler--DeWitt equation.

This makes us to look in a new light at the status of BRST
invariance. We have a theory in EPS which, after imposing a gauge
condition, is still invariant under global BRST transformation.
However, this simple example shows that after quantization of the
theory the requirement of BRST invariance is not, in general, a
remedy to restore the broken gauge invariance.

\section{The general solution to the Schr\"odinger equation and
the role of gauge degrees of freedom}
Let us turn now to the quantization procedure. Independently of our
notion of gauge invariance or noninvariance of the theory, the wave
function should obey some Schr\"odinger equation. We derive the
Schr\"odinger equation from the path integral with the effective
action (\ref{action}) and without asymptotic boundary conditions by
a standard method originated by Feynman \cite{Fey,Cheng}. So, the
Schr\"odinger equation is a direct mathematical consequence of the
fact that we consider the Universe {\it as a system without
asymptotic states}. For the present model it reads
\begin{equation}
\label{SE1}
i\,\frac{\partial\Psi(\mu,Q,\theta,\bar\theta;\,t)}{\partial t}
 =H\Psi(\mu,\,Q,\,\theta,\,\bar\theta;\,t),
\end{equation}
where
\begin{equation}
\label{H}
H=-\frac i w\frac{\partial}{\partial\theta}
   \frac{\partial}{\partial\bar\theta}
  -\frac1{2M}\frac{\partial}{\partial Q^{\alpha}}MG^{\alpha\beta}
   \frac{\partial}{\partial Q^{\beta}}+\frac1v(U-V);
\end{equation}
the operator $H$ corresponds to the Hamiltonian in EPS
(\ref{Hamilt}). $M$ is the measure in the path integral,
\begin{equation}
\label{M}
M(\mu, Q)=v^{\frac K2}(\mu, Q)w^{-1}(\mu, Q);
\end{equation}
$K$ is a number of physical degrees of freedom; the wave function is defined
on extended configurational space with the coordinates
$\mu,\,Q^a,\,\theta,\,\bar\theta$.
$V$ is a quantum correction to the potential $U$, that depends on the chosen
parametrization (\ref{paramet}) and gauge (\ref{frame_A}) (its
explicit form has been given, e. g. in \cite{Shest2,Shest3}).

The general solution to the Schr\"odinger equation has the following structure:
\begin{equation}
\label{GS-A}
\Psi(\mu,\,Q,\,\theta,\,\bar\theta;\,t)
 =\int\Psi_k(Q,\,t)\,\delta(\mu-f(Q)-k)\,(\bar\theta+i\theta)\,dk.
\end{equation}
It is a superposition of eigenstates of a gauge operator,
\begin{equation}
\label{k-vector}
\left(\mu-f(Q)\right)|k\rangle=k\,|k\rangle;\quad
|k\rangle=\delta\left(\mu-f(Q)-k\right).
\end{equation}
It can be interpreted in the spirit of Everett's ``relative state''
formulation. In fact, each element of the superposition
(\ref{GS-A}) describe a state in which the only gauge degree of
freedom $\mu$ is definite, so that time scale is determined by
processes in the physical subsystem through functions
$v(\mu,\,Q),\,f(Q)$ (see (\ref{paramet}), (\ref{frame_A})), while
$k$ being determined by initial clock setting. Indeed, according to
(\ref{frame_A}), the parameter $k$ gives an initial condition for
the variable $\mu$. The function $\Psi_k(Q,\,t)$ describes a state
of the physical subsystem for a reference frame fixed by the
condition (\ref{frame_A}). It is a solution to the equation
\begin{equation}
\label{phys.SE}
i\,\frac{\partial\Psi_k(Q;\,t)}{\partial t}
 =H_{(phys)}\Psi_k(Q;\,t),
\end{equation}
\begin{equation}
\label{phys.H-A}
H_{(phys)}=\left.\left[-\frac1{2M}\frac{\partial}{\partial Q^a}
  \frac1v M\gamma^{ab}\frac{\partial}{\partial Q^b}
 +\frac1v (U-V)\right]\right|_{\mu=f(Q)+k}.
\end{equation}

The dependence of $\Psi_k(Q,\,t)$ on $k$ is not fixed by the
equation (\ref{phys.SE}) in the sense that $\Psi_k(Q,\,t)$ can be
multiplied by an arbitrary function of $k$. On the other side, one
cannot choose the function $\Psi_k(Q,\,t)$ to be not depending on
$k$, since in this case one would obtain a non-normalizable,
non-physical state. The normalization condition for the wave
function (\ref{GS-A}) reads
$$
\int\Psi^*(\mu,\,Q,\,\theta,\,\bar\theta;\,t)\,
 \Psi(\mu,\,Q,\,\theta,\,\bar\theta;\,t)\,M(\mu,\,Q)\,
 d\mu\,d\theta\,d\bar\theta\,\prod_adQ^a=
$$
$$
\int\Psi^*_k(Q,\,t)\,\Psi_{k'}(Q,\,t)\,
 \delta(\mu-f(Q)-k)\,\delta(\mu-f(Q)-k')\,M(\mu,\,Q)\,
 dk\,dk'\,d\mu\,\prod_adQ^a=
$$
\begin{equation}
\label{Psi_norm}
=\int\Psi^*_k(Q,\,t)\,\Psi_k(Q,\,t)\,
 M(f(Q)+k,\,Q)\,dk\,\prod_adQ^a=1.
\end{equation}
It is normalizable under the condition that $\Psi_k(Q,\,t)$ is a
sufficiently narrow packet over $k$. Even from the classical point
of view a gauge condition cannot be fixed absolutely precisely, so
that the wave function (\ref{GS-A}) would depend on the gauge
conditions through $\delta$-function. We should rather consider a
narrow enough packet over $k$ to fit a certain classical $\bar k$
value:
\begin{eqnarray}
\Psi(\mu,Q,\theta,\bar\theta;\,t)&\!=\!&
 \frac1{\sqrt{2i\alpha\sqrt\pi}}\,
 \int\limits_{-\infty}^{\infty}
  \exp\left[-\frac1{2\alpha^2}\left(k-\bar k\right)^2\right]
  \Psi_{\bar k}(Q,\,t)\delta(\mu-f(Q)-k)
  \left(\bar\theta+i\theta\right)dk=\nonumber\\
\label{pack}
&\!=\!&\frac1{\sqrt{2i\alpha\sqrt\pi}}\,
 \exp\left[-\frac1{2\alpha^2}\left(\mu-f(Q)-\bar k\right)^2\right]\,
 \Psi_{\bar k}(Q,\,t)\,\left(\bar\theta+i\theta\right).
\end{eqnarray}

Our main aim is to give a description of a physical Universe, so
our next step will be the construction of a density matrix
\begin{equation}
\label{rho_general}
\rho(Q,Q',t)=
 \int\Psi^*(\mu,\,Q,\,\theta,\,\bar\theta;\,t)\,
  \Psi(\mu,\,Q',\,\theta,\,\bar\theta;\,t)\,M(\mu,\,Q)\,
  d\mu\,d\theta\,d\bar\theta.
\end{equation}
For the wave function (\ref{pack}) the expression for density
matrix reads
\begin{eqnarray}
\rho(Q,Q',t)&\!=\!&
 \exp\left(-\frac1{4\alpha^2}\left[f(Q)-f(Q')\right]^2\right)\,
 \Psi^*_{\bar k}(Q,\,t)\,\Psi_{\bar k}(Q',\,t)\times\nonumber\\
\label{rho_part}
&\!\times\!&M\left(\frac12\left[f(Q)+f(Q')\right]+\bar k,\,Q\right).
\end{eqnarray}
The normalization condition for the density matrix is
\begin{equation}
\label{rho_norm}
\int\rho(Q,Q,t)\prod_adQ^a=
 \int\Psi^*_{\bar k}(Q,\,t)\,\Psi_{\bar k}(Q,\,t)\,
  M\left(f(Q)+\bar k,\,Q\right)\prod_adQ^a=1,
\end{equation}
it corresponds to the condition (\ref{Psi_norm}).

By introducing a certain gauge condition we determine a gauge
subsystem of the Universe which affects properties of physical
Universe. From a mathematical point of view we can single out
three points where the gauge subsystem shows itself as a real
constituent of the Universe:
\begin{itemize}
\item A chosen gauge condition determines the form of the equation
(\ref{phys.SE}) for the physical part of the wave function
$\Psi_k(Q,\,t)$, in particular, an effective quantum potential.
\item The density matrix (\ref{rho_part}) describing physical
subsystem of the Universe also depends on the chosen gauge
condition through the factor
$\exp\left(-\frac1{4\alpha^2}\left[f(Q)-f(Q')\right]^2\right)$. It
plays an important role when analyzing the question under what
conditions the Universe can behave in a classical manner.
\item The measure $M\left(f(Q)+\bar k,\,Q\right)$ in physical
subspace also depends on the gauge condition, so that any changes
of the gauge condition result in changes of the measure. In other
words, if we determined the gauge subsystem in some different way,
it would reflect on the structure of physical subspace.
\end{itemize}

\section{Problems and prospects}
In this concluding part of my talk I shall consider in more detail
these three points.
\begin{itemize}
\item One can draw an analogy between solutions to the equation
(\ref{phys.SE}) for the physical part of the wave function
corresponding to different reference frames and solutions to the
Wheeler -- DeWitt equations in different parametrization mentioned
in the Introduction.

In general, one can seek a solution to Eq.(\ref{phys.SE}) in the
form of superposition of stationary state eigenfunctions:
\begin{equation}
\label{stat.WF}
\Psi_k(Q,\,t)=\sum_n c_n\Psi_{kn}(Q)\exp(-iE_n t);
\end{equation}
\begin{equation}
\label{stat.states}
H_{(phys)}\Psi_{kn}(Q)=E_n\Psi_{kn}(Q).
\end{equation}
The eigenvalue $E$ corresponds to a new integral of motion that
emerges as a result of fixing a gauge condition and characterizes
the gauge subsystem. On the other hand, in some papers
\cite{CFL,Fil,DF} it has been suggested to reduce the Wheeler --
DeWitt equation to a stationary Schr\"odinger equation of the form
$H\Psi=E\Psi$ (see also \cite{MM1,MM2,MM3}, where the canonical
formalism is modified in such a way that a new Hamiltonian
constraint becomes parabolic resulting in a Schr\"odinger equation
and corresponding eigenvalue problem for a wave function). The
origin of the Hamiltonian eigenvalue is different in different
approaches, but it would be interesting to compare a physical sense
of these solutions. In any case, there still remains the problem of
choice of parametrization, or, strictly speaking, a preferable
reference frame.

\item About 15 years ago a number of papers were published where
the notion of decoherence in quantum cosmology was widely
discussed. Decoherence is believed to be one of necessary
requirements for the system (here we mean the Universe as a whole)
could be regarded as classical \cite{Hal2}. It implies a transition
from a pure state to some mixed state described by diagonal density
matrix which could be considered as a result of interaction with
some environment \cite{Zurek, UZ}. The application of this idea to
quantum cosmology implies splitting the Universe into two
subsystems, one of which is a system under investigation and the
other plays the role of environment. It has been proposed to
consider some modes of scalar, gravitational and other fields as an
environment. However, there is no natural way to split the Universe
into two subsystems.

Our investigation \cite{Shest3} demonstrated that the gauge
subsystem can be supposed to be a candidate for the role of an
environment. Gauge degrees of freedom are not observable directly,
but only by its influence on the physical subsystem. The expression
(\ref{rho_part}) shows that under some gauge condition the density
matrix becomes about diagonal. A good example is given by the
condition for the lapse function $N$:
\begin{equation}
\label{N-cond}
N=a+\frac1{a^3}.
\end{equation}
(here $a$ is a scalar factor). This gauge condition is rather
interesting in some respects. For large $a$ we have $N=a$
(conformal time gauge), while in the limit of small $a$ the
condition (\ref{N-cond}) can be rewritten as $N a^3=1$. The latter
corresponds to the constraint on metric components $\sqrt{-g}={\rm
const}$. This constraint is known to lead to the appearance of
$\Lambda$-term in the Einstein equations \cite{Weinberg, Shest1}.
So the condition (\ref{N-cond}) describes in the limit of small $a$
an exponentially expanded early universe with $\Lambda$-term while
in the limit of large $a$ we have a Friedmann universe in the
conformal time gauge $N=a$. For the gauge (\ref{N-cond})
$f(a)=a+\frac1{a^3}$, and in the limit of large $a$ the density
matrix would have a Gaussian peak,
$\exp\left[-\frac1{4\alpha^2}(a-a')^2\right]$, i. e. the Universe
would demonstrate a classical behaviour in a good approximation.

In our approach the Universe may behave quasi-classically only in a
spacetime region where one can introduce a certain gauge
condition. If spacetime manifold consists of regions covered by
different coordinate charts, so that one should introduce different
reference frames in these regions, the Universe cannot behave in a
classical manner nearby borders of these regions. It is in
accordance with the fact that the quasiclassical approximation is
also not valid nearby borders of these regions. We can consider it
as another indication that the description of reference frames
should not be ignore when one deals with nontrivial topology of
spacetime.

\item As was said above, any changes of a gauge condition result
in changes of the measure $M\left(f(Q)+\bar k,\,Q\right)$ and, in
general, in the structure of physical subspace. Varying of the
gauge condition means that from the basis (\ref{k-vector}) one
goes to another basis \cite{Shest2},
\begin{equation}
\label{frame_C}
\left(\mu-f(Q)-\delta f(Q)\right)|k\rangle=k\,|k\rangle;\quad
|k\rangle=\delta\left(\mu-f(Q)-\delta f(Q)-k\right).
\end{equation}
The Hamiltonian in physical subspace (\ref{phys.H-A}) whose form
also depends on chosen parametri\-zation and gauge then could acquire
additional terms, so that with respect to the original subspace
with the basis (\ref{k-vector}) the Hamiltonian may, in general,
acquire an anti-Hermitian part. It would means that the transition
to another basis (\ref{frame_C}) would have an irreversible
character from the point of view of the observer in the reference
frame corresponding to the basis (\ref{k-vector}). On the other
side, the observer in the reference frame corresponding to the
basis (\ref{frame_C}) would not experience any irreversible changes
and, from his point of view, the evolution of the Universe would be
unitary. There is a certain analogy between the transition from one
to another basis in physical subspace and a transition to
accelerating reference frame in Rindler space. In the latter case
one can formally associate some entropy with the area of an
accelerating horizon \cite{HH1,JP}. However, for the observer in a
non-accelerating reference frame the entropy would be zero and time
evolution would be described by an unitary operator. In the both
cases we can see that physical situations are observer dependent.
But, in the case of nontrivial topology which was mentioned above,
we {\it have to pass on to another reference frame and another
basis in physical subspace}. It rather resembles the situation with
casual horizons when the very existence of black hole entropy poses
the question if evolution is still unitary.

Further investigation of these questions requires development of
the formalism beyond simple models with finite degrees of freedom.
The generalization of the formalism for time dependent gauges, when
the structure of physical subspace is continuously changing, also
deserves careful attention. It is another prospect of the presented
approach.
\end{itemize}

\section*{Acknowledgments}
The author is grateful to the Organizing Committee of
``COSMION-2004'' for financial support which let her participate
in the conference.

\end{document}